\documentclass{PoS}
\title{Studies of the internal properties of jets and jet substructure with the ATLAS Detector}
\ShortTitle{Jet substructure with ATLAS}

\usepackage{xspace}
\usepackage{subfigure}
\usepackage{graphicx} 
\usepackage{url}
\usepackage{cite}

\input{atlasphysics.sty}
\newcommand{\pp}        {\ensuremath{pp}\xspace}
\newcommand{\seventev}  {\ensuremath{7~\mathrm{TeV}}\xspace}
\newcommand{\sqsseven}  {\ensuremath{\sqrt{s} = \seventev}\xspace}
\newcommand{\invpb}     {~\ensuremath{{\rm pb}^{-1}}\xspace}
\newcommand{\kt}        {\ensuremath{k_{t}}\xspace}
\newcommand{\AKT}       {anti-\kt}
\newcommand{\antikt}    {\AKT}

\newcommand{\AKTFat}    {\AKT, \ensuremath{R=1.0}\xspace}
\newcommand{\CamKt}     {\ensuremath{\mathrm{C/A}}\xspace}

\newcommand{\CAFat}     {\CamKt, \ensuremath{R=1.2}\xspace}

\newcommand{\ytwoth}    {\ensuremath{y_{23}}\xspace}

\newcommand{\mjet}      {\ensuremath{m^{\rm jet}}\xspace}

\newcommand{\dip}       {\ensuremath{\mathcal{D}}\xspace}

\newcommand{\DOneTwo}   {\ensuremath{\sqrt{d_{12}}}\xspace}
\newcommand{\DTwoThr}   {\ensuremath{\sqrt{d_{23}}}\xspace}

\newcommand{\Npv}       {\ensuremath{N_{\rm PV}}\xspace}

\newcommand{\Npp}       {\ensuremath{N_{pp}}\xspace}

\newcommand{\lumi}      {\ensuremath{\mathcal L}\xspace}

\newcommand{\Pythia}    {\texttt{PYTHIA}\xspace}
\newcommand{\Herwig}    {\texttt{HERWIG}\xspace}
\newcommand{\herwigpp}  {\texttt{HERWIG}++\xspace}
\newcommand{\Herwigpp}  {\herwigpp}
\newcommand{\Jimmy}     {\texttt{JIMMY}\xspace}
\newcommand{\Alpgen}    {\texttt{ALPGEN}\xspace}
\newcommand{\HJ}        {\Herwig\!/\Jimmy\xspace}

\newcommand{\MCATNLO}   {\textsc{MC@NLO}\xspace}

\newcommand{\comment}[1]{}

\author{\speaker{David W. Miller}
        \thanks{Now at The University Of Chicago, Enrico Fermi Institute.}\\
        On behalf of the ATLAS Collaboration\\
        SLAC National Accelerator Laboratory\\
        E-mail: \email{David.W.Miller@uchicago.edu}}

\abstract{
The internal structure of jets produced in \pp collisions at the LHC is measured using the ATLAS detector in an inclusive jet sample corresponding to 35\invpb of \pp collisions at \sqsseven. Classical jet shape and energy flow measurements are complemented with measurements of new substructure observables with comparisons made to several leading order parton shower Monte Carlo programs. The jet invariant mass and \kt splitting scale are measured for \AKT jets with a distance parameter of $R=1.0$ and Cambridge-Aachen jets with $R=1.2$. Furthermore, a splitting and filtering procedure is applied to the Cambridge-Aachen jets. These tools are then utilized for the first measurements of the filtered jet mass at the LHC in the inclusive jet sample as well the $W$+1 jet sample, in which a hadronic $W$ mass peak is observed in the jet invariant mass spectrum. A sample of candidate boosted top quark events is also analyzed in detail for the jet substructure properties of hadronic ``top-jets'' in the final state.
}

\FullConference{The 2011 Europhysics Conference on High Energy
Physics-HEP 2011,\\
July 21-27, 2011\\
Grenoble, Rh\^one-Alpes France}

\begin{document}


\section{Introduction}
\label{sec:intro}
The internal structure of individual jets in ATLAS~\cite{detPaper} is extended beyond classical~\cite{Collaboration:2011kq, ATL-PHYS-PUB-2011-010} jet shapes with measurements of the jet invariant mass and the \kt splitting scales. These studies are performed in the context of understanding QCD and the potential for hadronic heavy particle decays such as $W$ bosons~\cite{Seymour:1993mx, Butterworth:2002tt, cscnote}, top quarks~\cite{ATL-PHYS-PUB-2010-008, brooijmans2, brooijmans, Chekanov:2010vc, Chekanov:2010gv}, the Higgs boson~\cite{Butterworth:2008iy, ATLASHV}, and potential new particles~\cite{Butterworth:2009qa}, to be collimated into a single heavy jet characterized by distinct substructure and large mass.  Two ``fat'' jet algorithms are used, along with the splitting and filtering jet grooming technique~\cite{Butterworth:2008iy, ATLASHV}. The data are corrected for detector effects and compared to the predictions from several Monte Carlo simulations implementing leading-order perturbative QCD matrix elements supplemented with
parton showers.  A first measurement of the jet invariant mass in ATLAS is made, and an additional jet substructure observable, the first \kt splitting scale or \DOneTwo, is measured for the first time at the LHC. A sample of candidate boosted top quark events exhibits the expected jet substructure properties and a hadronic $W$ mass peak is observed in the jet invariant mass spectrum.

\section{Data selection and Monte Carlo samples}
\label{sec:data}
Events collected by the ATLAS detector are selected to have a high-\pT\ jet ($\pT > 300$~GeV) in the final state using a fully efficient inclusive jet trigger corresponding to an integrated luminosity of approximately 35\invpb. The jet mass and \kt\ splitting scales are measured for jets reconstructed by the \AKT~\cite{Cacciari:2008gp, Cacciari200657} algorithm with a distance parameter $R=1.0$ and \CAFat~\cite{Dokshitzer:1997in,Wobisch:1998wt}. Jet quality criteria reject events contaminated by beam-related backgrounds or detector defects.

The data are compared to events simulated using several Monte Carlo (MC) programs that are also re-weighted to match the number of multiple proton-proton interactions observed.  \Alpgen 2.13~\cite{alpgen} is interfaced to \Herwig~6.510~\cite{Corcella2001} for the parton shower and hadronization models and to \Jimmy~4.31~\cite{jimmy} for the underlying event model. Exact LO pQCD matrix elements with up to 6 partons in the final state are used by \Alpgen. The LO MC programs (for inclusive jet production) \Pythia~6.423~\cite{Sjostrand2001} and \Herwigpp~2.4~\cite{Bahr2008} rely on the parton shower to produce the equivalent of multi-parton final states. \Pythia and \Herwigpp provide shower models which are $p^{2}_T$-ordered and angular-ordered, respectively. Comparisons of the jet mass in Section~\ref{sec:results} are also made to \Herwig interfaced to \Jimmy for the underlying event model.


\section{Jet substructure}
\label{sec:results}
The \KT\ splitting scale, $\sqrt{d_{ij}}$, is defined by reclustering a jet with the \KT\
algorithm~\cite{kt,kt2} such that $\sqrt{d_{ij}}=\mathrm{min}(p_{T,i},p_{T,j}) \times \delta R_{i,j}$, where $\delta R_{i,j} = \sqrt{d\phi_{i,j}^2 + dy_{i,j}^2}$ and $i,j$ represent the last two proto-jets in the recombination~\cite{ATLAS-CONF-2011-073}. For \CamKt jets, the filtering procedure identifies relatively hard, symmetric splittings in a jet that contribute significantly to the jet invariant mass~\cite{ATLAS-CONF-2011-073}.

The jet mass and splitting scale are corrected for detector effects via a bin-by-bin unfolding procedure. Bin sizes are chosen so that bin migrations are small and the per-bin purity >50\%.

\begin{figure}
  \centering
  \subfigure[]{
    \includegraphics[width=0.35\columnwidth]{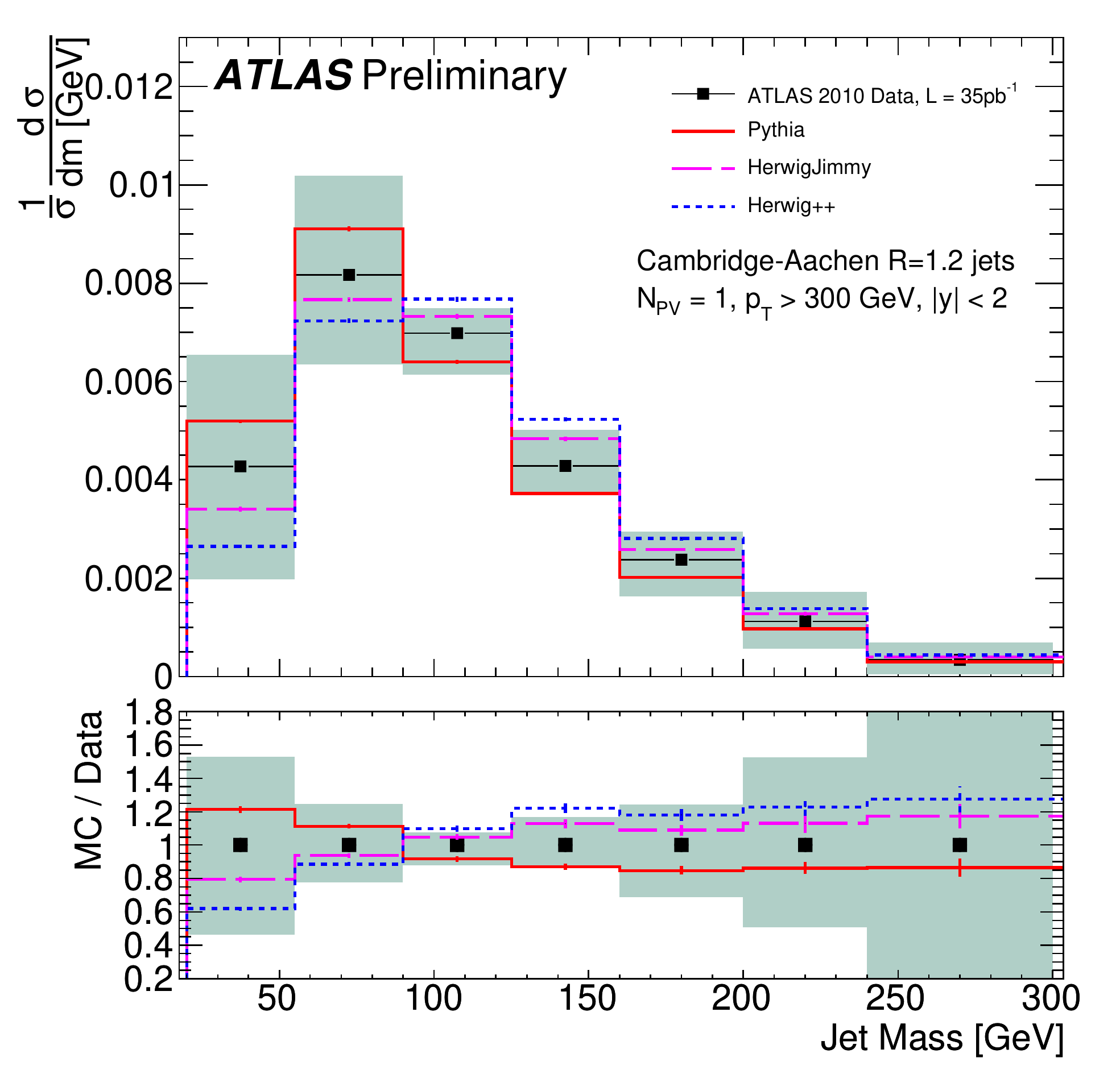}
    \label{fig:results:camass:unfiltered}}
  \subfigure[]{
    \includegraphics[width=0.35\columnwidth]{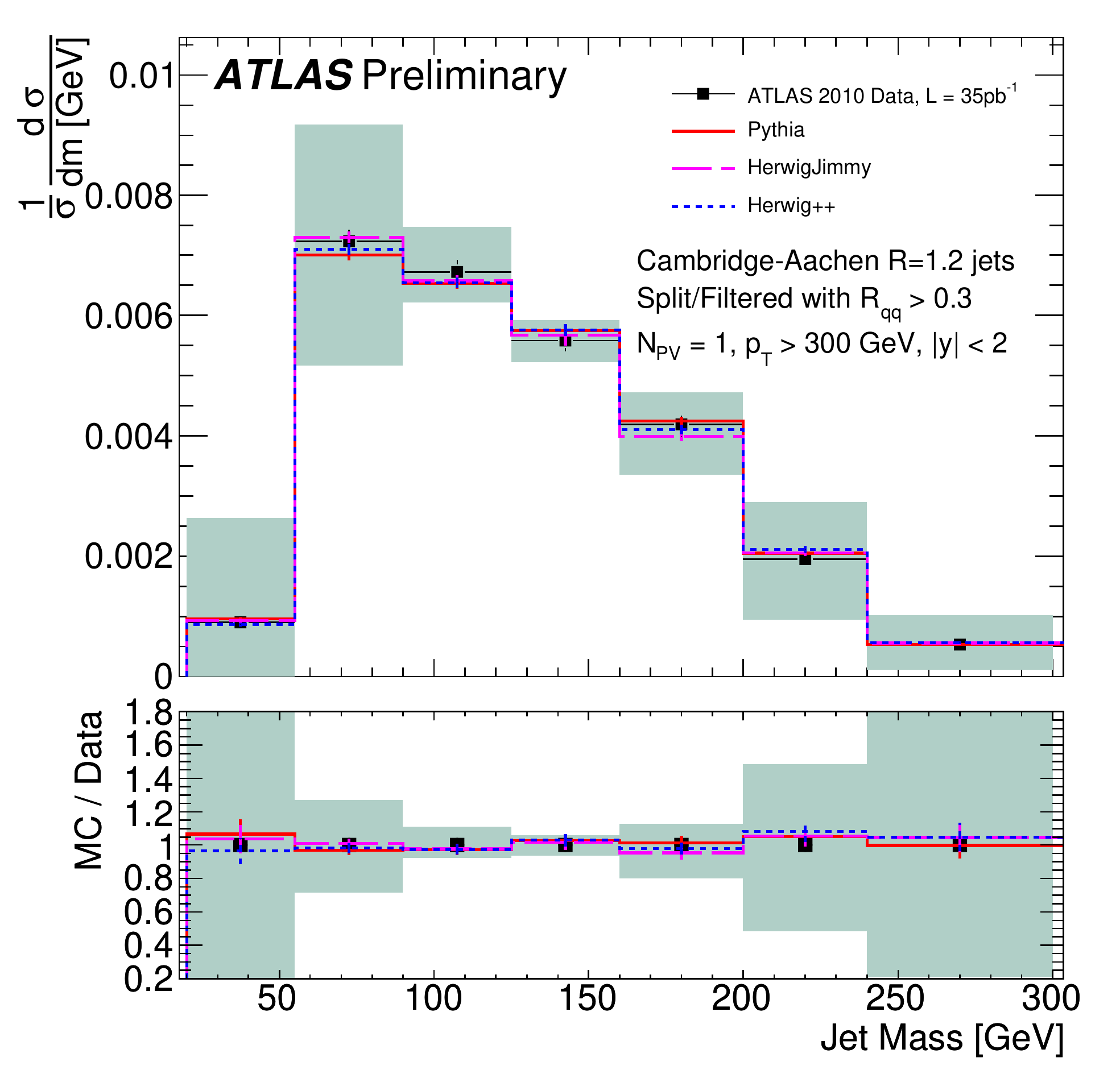}
    \label{fig:results:camass:filtered}}
  \subfigure[]{
    \includegraphics[width=0.35\columnwidth]{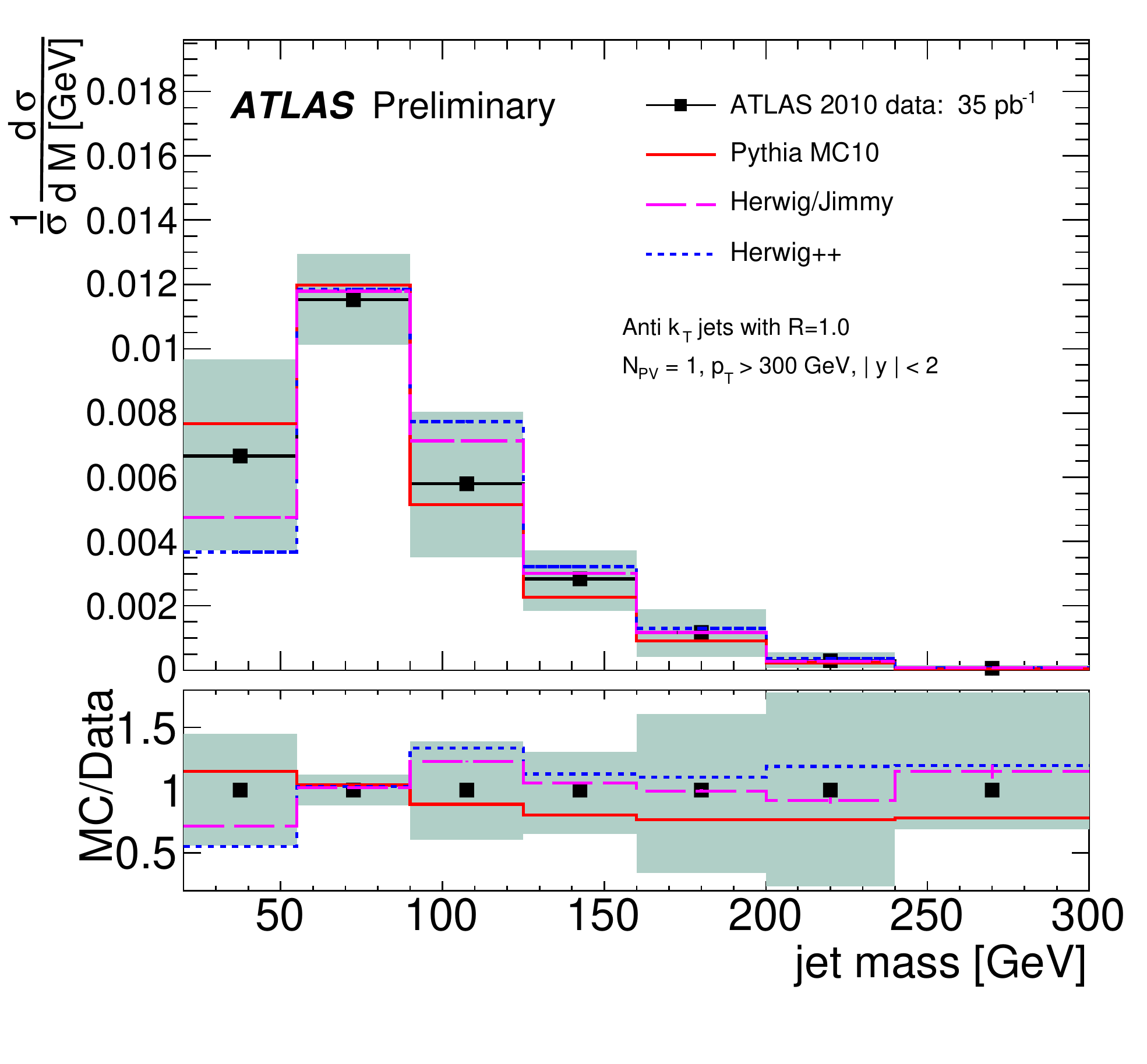}
    \label{fig:results:aktmass:mass}}
  \subfigure[]{
    \includegraphics[width=0.35\columnwidth]{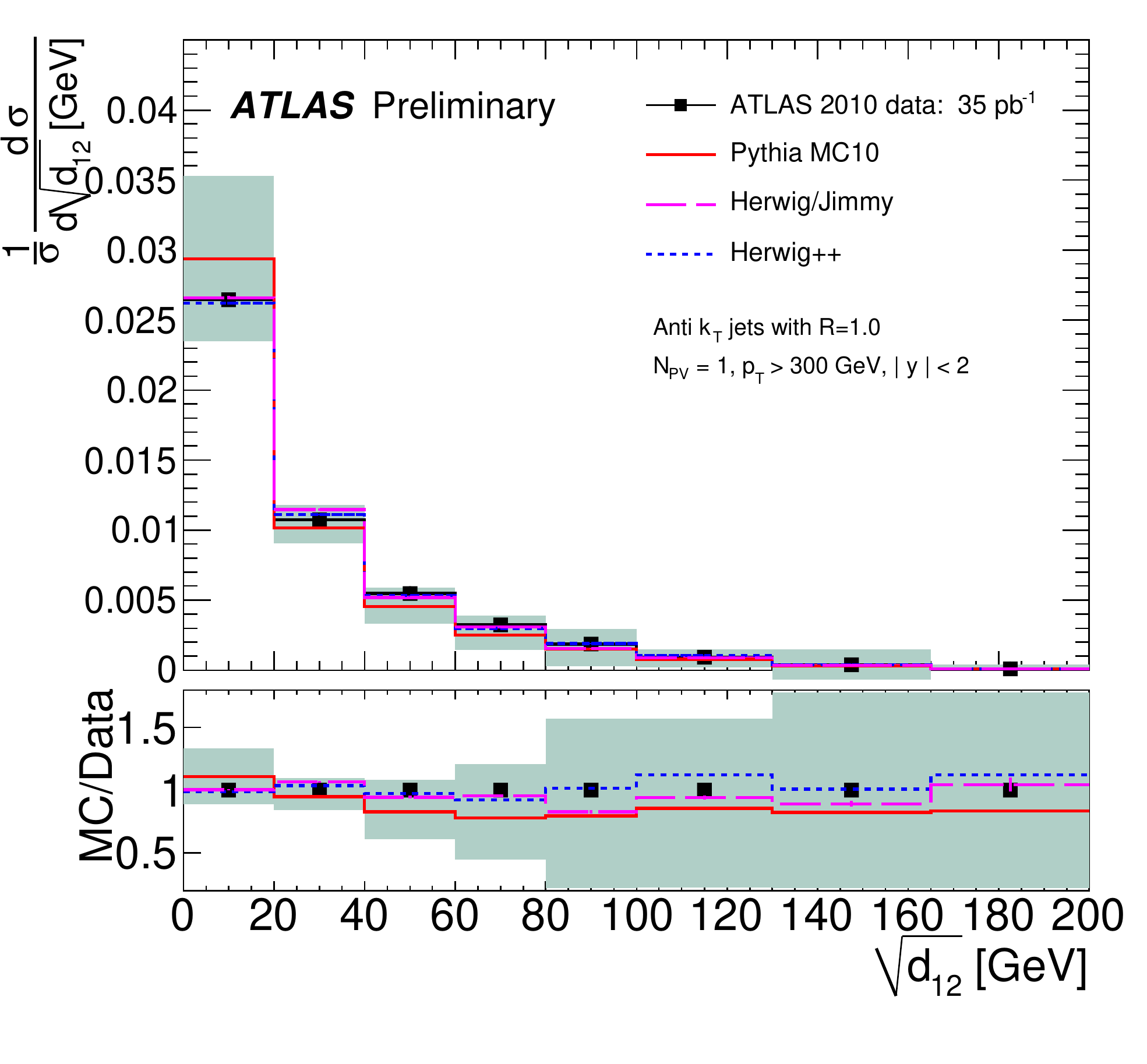}
    \label{fig:results:aktmass:d12}}
  \caption{
    Invariant mass spectrum of 
    (a) Cambridge-Aachen jets before and 
    (b) after the filtering procedure has 
    been applied. 
    For \antikt jets,
    (c) the invariant mass spectrum and the
    (d) \DOneTwo spectrum for the same jets.
    All distributions are fully corrected
    for detector effects, systematic uncertainties are depicted by the
    shaded band~\cite{ATLAS-CONF-2011-073}.
  }
  \label{fig:results}
\end{figure}

Fully corrected hadron-level distributions of the jet mass and \DOneTwo are shown in Figure~\ref{fig:results}~\cite{ATLAS-CONF-2011-073}. \Herwigpp tends to predict slightly more massive \CamKt jets than supported by the data whereas \Pythia and \Herwig/\Jimmy yield measurements which bracket the data and agree to within systematic uncertainties. Notably, the differences between the mass distributions predicted by the various MC programs are greatly reduced after the filtering procedure. This observation suggests that the jet mass after the filtering procedure accurately represents the true hard components within the jets which are modeled well by the simulations.

A similar conclusion may be drawn in the case of \AKTFat jets as for the \CamKt jets, wherein \Herwigpp tends to predict slightly more massive jets than observed in the data. In the case of \AKT, the high mass tail of the distribution is better reflected in the \Herwig/\Jimmy and \Herwigpp MC simulations than in \Pythia. 

Studies show~\cite{ATLAS-CONF-2011-073} that the average jet mass grows linearly with $R$ in the case of zero additional $pp$ interactions (pile-up), whereas the slope as a function of pile-up shows an $R^3$ dependence. The filtering procedure removes this dependence to within statistical and systematic uncertainties.

\section{Boosted heavy particles}
\label{sec:particles}
\begin{figure}
  \centering
    \includegraphics[width=0.45\columnwidth]{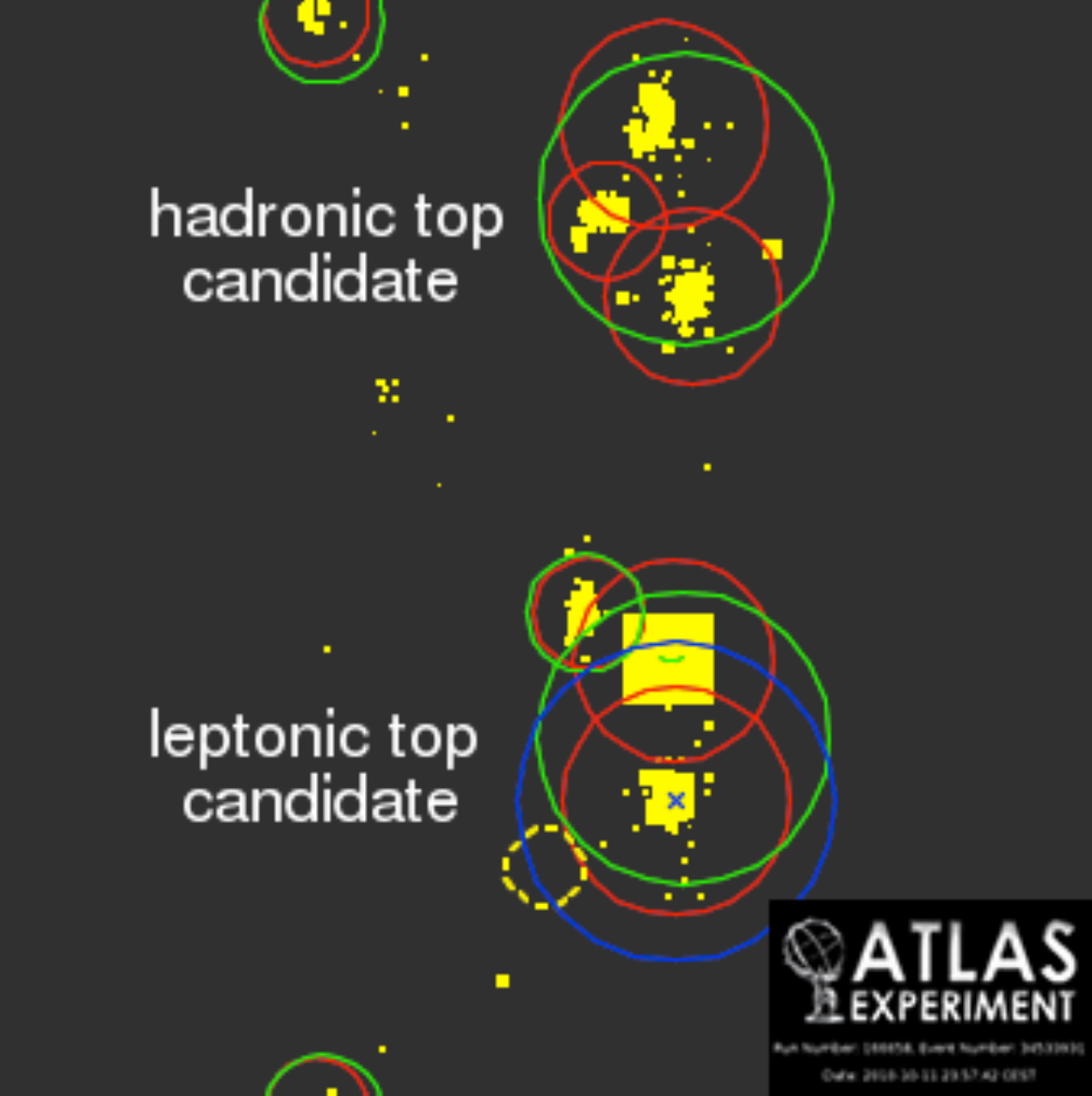}
    \includegraphics[width=0.48\columnwidth]{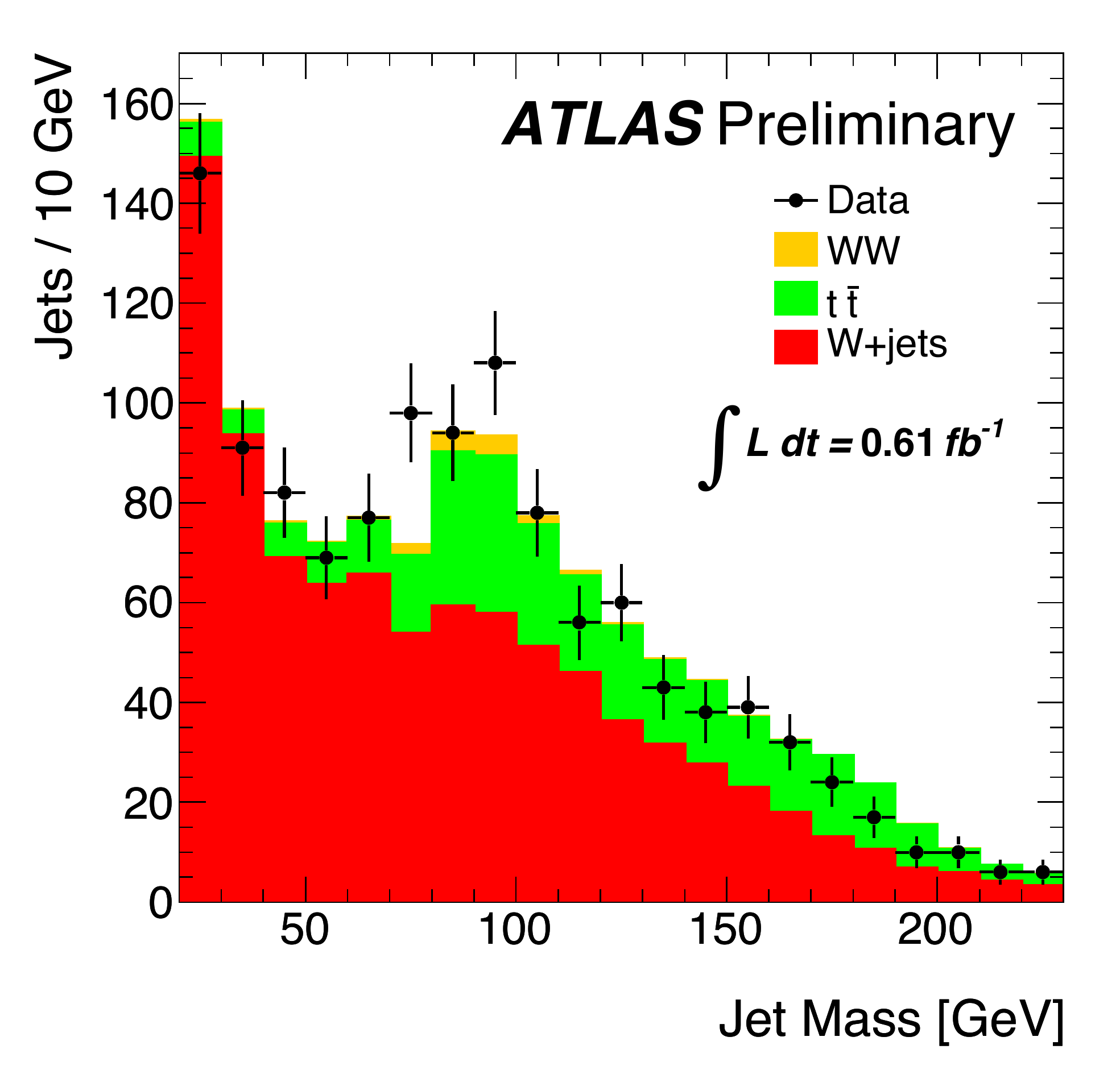}
  \caption{
    (left) Candidate boosted top quark candidate. Details of this event are described in Ref.~\cite{ATLAS-CONF-2011-073}.
    (right) Jet invariant mass distribution in $W$ + 1 jet events~\cite{ATLAS-CONF-2011-103}. 
  }
  \label{fig:particles}
\end{figure}

The use of jet substructure is tested in a small sample of candidate boosted top quark and hadronically decaying $W$ bosons in a $W$+1 jet sample. Fully reconstructed top quark pair events are selected in the lepton+jets channel to have $m_{\ttbar}>700~\GeV$. Figure~\ref{fig:particles} shows one of the selected events wherein the \AKTFat jet corresponding to the hadronic top quark decay has $\pT = 327~\GeV$ and $\mjet=206~\GeV$ as well as $\DOneTwo=110$~GeV and $\DTwoThr=40$~GeV. 

Candidate hadronic $W$ events are selected to contain a $W\to\ell\nu$ candidate with $\pt^{W} > 200 \GeV$~\cite{ATLAS-CONF-2011-103}. The jet mass distribution of filtered \CAFat jets with $\pt > 180 \GeV$ and $\Delta \phi_{W,{\rm jet}} > 1.2$ in these events is shown in Figure~\ref{fig:particles}. The three main contributions to these events are \ttbar\ (generated with \MCATNLO+\Herwig/\Jimmy~\cite{mcatnlo,mcatnlo2}), $W$+jets (generated with \Alpgen+\Herwig/\Jimmy),  and $WW$ (generated with \Herwig/\Jimmy), all normalized to the highest order cross-section available (see Ref.~\cite{ATLAS-CONF-2011-103} for more details). The good agreement between data and the various MC simulations suggests both that the tools described above are well described in a complex physics environment and that the systematics are generally well under control.

\section{Conclusions}
\label{sec:conclusions}
The substructure of hadronic jets is studied in terms of the jet mass and \kt\ splitting scales for \AKTFat and \CAFat jets, as well as \CamKt jets with filtering applied. In all observables the \Pythia and \Herwig samples are in agreement with data to within the systematic uncertainties. The \Herwigpp prediction appears to be slightly disfavoured in the unfiltered \CamKt mass spectra, producing jets with a higher mass than found in data.

Overall it is clear that ATLAS is capable of delivering measurements of the variables considered in this study and that these observables are well modeled by leading order Monte Carlo. Early applications of these techniques are already demonstrated through the tagging of candidate boosted top quarks and the observation of fully hadronic $W$ decays. It is expected that searches for boosted Higgs bosons, supersymmetric particles, and top-quark resonances will benefit from such advanced techniques.


\bibliographystyle{unsrtstanford}
\bibliography{substructure}

\end{document}